\documentclass[12pt]{article}
\newcommand{\nc}{\newcommand}
\nc{\fh}{\hat{f}}
\nc{\muh}{\hat{\mu}}
\nc{\nuh}{\hat{\nu}}
\nc{\bib}{\bibitem}
\nc{\al}{\alpha}
\nc{\g}{\gamma}
\nc{\G}{\Gamma}
\nc{\D}{\Delta}
\nc{\eps}{\epsilon}
\nc{\la}{\lambda}
\nc{\La}{\Lambda}
\nc{\var}{\varphi}
\nc{\cg}{{\cal G}}
\nc{\pa}{\partial}
\nc{\nn}{\nonumber \\ }
\nc{\hf}{\frac{1}{2}}  
\nc{\dz}{\frac{dz}{2\pi i}}
\nc{\bin}[2]{\left (\begin{array}{c} {#1}\\ {#2} \end{array}\right )}
\nc{\ben}{\begin{equation}}
\nc{\een}{\end{equation}}
\nc{\bea}{\begin{eqnarray}}
\nc{\eea}{\end{eqnarray}}
\nc{\bra}[1]{\langle {#1}|}
\nc{\ket}[1]{|{#1}\rangle}
\newcommand{\Z}{\mbox{$Z\hspace{-2mm}Z$}}
\nc{\C}{\mbox{\hspace{1.24mm}\rule{0.2mm}{2.5mm}\hspace{-2.7mm} C}}
\nc{\Nat}{\mbox{\hspace{.04mm}\rule{0.2mm}{2.8mm}\hspace{-1.5mm} N}}
\nc{\HH}{\mbox{\hspace{.04mm}\rule{0.2mm}{2.8mm}\hspace{-1.5mm} H}}

\def\vvdots{\mathinner{\mkern1mu\raise1pt\vbox{\kern7pt\hbox{.}}\mkern2mu
 \raise4pt\hbox{.}\mkern2mu\raise7pt\hbox{.}\mkern1mu}}

\begin{document}

\topmargin -5mm
\oddsidemargin 5mm

\begin{titlepage}
\setcounter{page}{0}

\vspace{8mm}
\begin{center}
{\huge Logarithmic limits of minimal models}

\vspace{15mm}
{\Large J{\o}rgen Rasmussen}\\[.3cm] 
{\em Centre de recherches math\'ematiques, Universit\'e de Montr\'eal}\\ 
{\em Case postale 6128, 
succursale centre-ville, Montr\'eal, Qc, Canada H3C 3J7}\\[.3cm]
rasmusse@crm.umontreal.ca

\end{center}

\vspace{10mm}
\centerline{{\bf{Abstract}}}
\vskip.4cm
\noindent
It is discussed how a
limiting procedure of (super)conformal field theories may result
in logarithmic (super)conformal field theories. The construction
is illustrated by logarithmic limits of 
(unitary) minimal models in conformal field
theory and in $N=1$ superconformal field theory.
\\[.5cm]
{\bf Keywords:} Logarithmic conformal field theory,
$N=1$ superconformal field theory, minimal models. 
\end{titlepage}
\newpage
\renewcommand{\thefootnote}{\arabic{footnote}}
\setcounter{footnote}{0}

\section{Introduction}

It has long been speculated that certain limits of minimal models
in conformal field theory (CFT) may correspond to 
logarithmic CFTs (LCFTs). We refer to \cite{FMS} for a survey
on CFT and to \cite{Flo,Gab,Nic} for recent reviews of LCFT.
The first systematic study of LCFT appeared in \cite{Gur}.
Flohr, in particular, has discussed \cite{Flo-9605} how LCFTs appear 
at the 'boundary' of the set of minimal models ${\cal M}(p,p')$ 
by considering $p$ and $p'$ not coprime, where the minimal
models may be characterized by a pair of coprime integers $p>p'>1$.
The set of models ${\cal M}(p,p')$ with $p,p'\geq1$
is obviously discrete. We suggest to say that any such model, which is 
not a minimal model, belongs to the boundary of the set
of minimal models.

The objective here is to discuss how LCFT may be obtained
by a limiting procedure different from the one used in \cite{Flo-9605},
to which it does not seem to be directly related.
Our approach is quite general in its own right, and is illustrated by 
logarithmic limits of unitary or non-unitary minimal models in CFT
as well as in $N=1$ superconformal field theory (SCFT) 
\cite{Eic,BKT,FQS,Fri}. We shall present a general prescription
for constructing (super)conformal
Jordan cells, thereby rendering
the associated models logarithmic (super)conformal field theories.

The idea of our construction is to consider a sequence of 
conformal models labeled by an integer $n$, with focus
on a pair of primary fields in each conformal model appearing
in the sequence. To get a firmer grip on this, we introduce sequences of
primary fields and organize the former in equivalence classes. 
For finite $n$, the two fields must have different conformal
weights, while the weights of the associated sequences 
converge to the same (finite)
conformal weight, $\D$, as $n$ approaches infinity.
A Jordan-cell structure emerges if one considers a
particular linear and (for finite $n$) invertible map
of the two fields (or of the associated sequences)
into two new fields. Since the
original fields have different conformal weights, the new
fields do not both have well-defined conformal weights.
In the limit $n\rightarrow\infty$, the linear map is singular 
and thus not invertible (thereby mimicking the In\"on\"u-Wigner or
Saletan contractions known from the theory of Lie algebras), 
while the new set of fields
make up a Jordan cell of conformal weight $\D$.
The two-point functions of the new fields are also discussed.

A naive study of the representations of the fields in
two sequences of primary fields as $n\rightarrow\infty$
would in general not allow one to distinguish between the two representations
of the resulting pair of fields. The singular map mentioned above
ensures such a distinction. 
A merit of our construction is thus that it makes manifest that
certain representations remain different in the limit instead
of potentially producing multiple copies of a single representation.

The replica approach to systems with disorder is based on
techniques resembling the ones employed in the present work.
In \cite{CKT, Car}, for example, logarithmic divergencies of
correlators were obtained in the so-called replica limit where a 
certain parameter vanishes. The paradigm differs from ours, though,
as we are considering infinite sequences of conformal models.

The general construction of a LCFT as a limiting procedure
of CFTs is outlined in Section 2
and subsequently illustrated by logarithmic limits
of minimal models. This is extended to
SCFT and the $N=1$ superconformal minimal models
in Section 3. Section 4 contains some concluding remarks.

\section{Logarithmic limits of CFT}

\subsection{General construction}

A Jordan cell (of rank two) consists of two fields:
a primary field, $\Phi$, of conformal weight $\D$, and
its non-primary partner, $\Psi$. With a conventional relative normalization
of the fields, we have
\bea
 T(z)\Phi(w)&=&\frac{\D\Phi(w)}{(z-w)^2}+\frac{\pa\Phi(w)}{z-w}\nn
 T(z)\Psi(w)&=&\frac{\D\Psi(w)+\Phi(w)}{(z-w)^2}
   +\frac{\pa\Psi(w)}{z-w}
\label{T}
\eea
where $T$ is the Virasoro generator.
The associated two-point functions are known to be of the form
\ben
 \langle\Phi(z)\Phi(w)\rangle\ =\ 0,\ \ \ \ 
 \langle\Phi(z)\Psi(w)\rangle\ =\ \frac{A}{(z-w)^{2\D}},\ \ \ \ 
 \langle\Psi(z)\Psi(w)\rangle\ =\ \frac{B-2A\ln(z-w)}{(z-w)^{2\D}}
\label{two}
\een
with structure constants $A$ and $B$. Our goal is to construct
such a system in the limit of a sequence of ordinary CFTs.

To this end, let us consider a sequence of conformal models $M_n$, 
$n\in\Z_>$, with central charges, $c_n$, converging to the finite value
\ben
 \lim_{n\rightarrow\infty}c_n\ =\ c
\label{c}
\een
This appears to be a necessary condition for the limit of the
sequence to exist.
It is assumed that $M_n$ contains a pair of primary fields, $\var_n$ and
$\psi_n$, of conformal weights $\D+a_n$ and $\D+b_n$, respectively,
where
\ben
 \lim_{n\rightarrow\infty}a_n\ =\ \lim_{n\rightarrow\infty}b_n\ =\ 0
\label{ab}
\een
We are thus considering sequences of primary fields such as
$(\var_1,\var_2,...)$
where the element $\var_n$ belongs to $M_n$.
Their two-point functions are assumed to be of the form
\bea
 \langle\var_n(z)\var_n(w)\rangle&=&\frac{C_\var}{(z-w)^{2(\D+a_n)}}\nn
 \langle\psi_n(z)\psi_n(w)\rangle&=&\frac{C_\psi}{(z-w)^{2(\D+b_n)}}\nn
  \langle\var_n(z)\psi_n(w)\rangle&=&0
\label{phivar}
\eea
where, for simplicity, 
the fields have been normalized so that the non-vanishing
structure constants, $C_\var$ and $C_\psi$, are {\em independent} of $n$.
The conformal models $M_n$ could have an extended symmetry
in which case the fields would be characterized by additional
quantum numbers. Since our construction is essentially independent of such
an eventuality, we may a priori allow $a_n=b_n$ while assuming 
$\langle\var_n\psi_n\rangle=0$.
We find, though, that consistency requires $a_n\neq b_n$
after all.

We now introduce the linear and invertible map
\ben
 \left(\begin{array}{ll} \Phi_n\\ \Psi_n\end{array}\right)\ =\
  \left(\begin{array}{ll} \al_n&\beta_n\\ \g_n&\delta_n
    \end{array}\right)
     \left(\begin{array}{ll} \var_n\\ \psi_n\end{array}\right)  
\label{lin}
\een
and define 
\ben
 \Phi:=\ \lim_{n\rightarrow\infty}\Phi_n,\ \ \ \ \ \ \ 
 \Psi:=\ \lim_{n\rightarrow\infty}\Psi_n
\label{PhiPsi}
\een
In the picture with sequences of primary fields alluded to above,
we are thereby defining two new sequences with potentially different
properties as $n\rightarrow\infty$. The explicit behaviour of (\ref{PhiPsi}) 
depends, of course, on
the map (\ref{lin}). It should be emphasized that the resulting model
may not share the same standards and properties as the model
obtained by considering the original sequences in the limit 
$n\rightarrow\infty$. A similar problem is known from Lie algebra
contractions where the resulting algebra only satisfies
the Jacobi identities under certain conditions.

For $\Phi$ and $\Psi$ to constitute a Jordan pair,
we should consider a map which becomes {\em singular}
as $n$ approaches infinity.
We find that the simplest map meeting our needs is given by
\bea
 \al_n&=&\sqrt{\frac{(a_n-b_n)A}{C_\var}},\ \ \ \ \ \ \ \ \beta_n\ =\ 0\nn
 \g_n&=&\sqrt{\frac{A}{(a_n-b_n)C_\var}},\ \ \ \ \ \ \ \delta_n\ =\ 
  \sqrt{\frac{(a_n-b_n)B-A}{(a_n-b_n)C_\psi}}
\label{n}
\eea
for which it is evident that $a_n\neq b_n$. For example, we have
\bea
 T(z)\Psi(w)&=&\lim_{n\rightarrow\infty}\left\{T(z)\Psi_n(w)\right\}\nn
 &=&\lim_{n\rightarrow\infty}\left\{\frac{\left(\D+
   \frac{\al_n\delta_n b_n-\beta_n\g_n a_n}{\al_n\delta_n-\beta_n\g_n}\right)
   \Psi_n(w)
   +\frac{\g_n\delta_n(a_n-b_n)}{\al_n\delta_n-\beta_n\g_n}\Phi_n(w)}{
   (z-w)^2}+\frac{\pa_w\Psi_n(w)}{z-w}\right\}\nn
  &=&\frac{\D\Psi(w)+\Phi(w)}{(z-w)^2}
   +\frac{\pa\Psi(w)}{z-w}
\label{TPsin}
\eea
where the inverse map has been used in the rewriting.
The map (\ref{n}) has been chosen in order to reproduce the
two-point functions (\ref{two}) with the structure constants
given there. 
That this is satisfied follows from the expansion
\ben
 x^\eps\ =\ e^{\eps\ln(x)}\ =\ 1+\eps\ln(x)+\hf\eps^2\ln^2(x)+{\cal O}(\eps^3)
\label{x}
\een
needed when evaluating expressions like (\ref{phivar})
in the limit $n\rightarrow\infty$. For example, we have
\bea
 \langle\Psi(z)\Psi(w)\rangle&=&\lim_{n\rightarrow\infty}\left\{
  \langle\Psi_n(z)\Psi_n(w)\rangle\right\}\nn
 &=&\lim_{n\rightarrow\infty}\left\{ \frac{\g_n^2C_\phi}{(z-w)^{2(\D+a_n)}}+ 
   \frac{\delta_n^2C_\psi}{(z-w)^{2(\D+b_n)}}\right\}\nn
 &=&\frac{B-2A\ln(z-w)}{(z-w)^{2\D}}
\label{Psin}
\eea
It follows in this way that the fields 
(\ref{PhiPsi}) constitute a Jordan cell of conformal weight $\D$.
   
We wish to point out that a Jordan cell also emerges in the
special case when
either $a_n$ or $b_n$ (but not both) is zero for all $n$.
An example may be found in \cite{BS} where a so-called correlated
limit of the parafermionic CFT is discussed. It has been found that
the construction in \cite{BS} extends to the graded parafermions as well
\cite{Rabat}.

\subsection{Logarithmic limits of minimal models}

Here we illustrate the general construction above by 
considering limits of minimal models. The minimal model
${\cal M}(p,p')$ is characterized by the coprime integers $p$
and $p'$ which may be chosen to satisfy $p>p'>1$, without loss of generality. 
The central charge is given by
\ben
 c\ =\ 1-6\frac{(p-p')^2}{pp'}
\label{cmm}
\een
whereas the primary fields, $\phi_{r,s}$, have conformal weights given by
\ben
 \D_{r,s}\ =\ \frac{(rp-sp')^2-(p-p')^2}{4pp'}, \ \ \ \ \ \ \ \ \ 
   1\leq r<p', \ \ \ 1\leq s< p
\label{Dmm}
\een
The bounds on $r$ and $s$ define the Kac table of admissible
primary fields. With the identification 
\ben
 \phi_{r,s}\ =\ \phi_{p'-r,p-s}
\label{iden}
\een
there are $(p-1)(p'-1)/2$ distinct primary fields in the model.
These models are unitary provided $p=p'+1$, in which case
the further constraint $s\leq r$ takes into consideration
the identification (\ref{iden}).

For each positive integer $k$ we now consider the sequence of minimal
models ${\cal M}(kn+1,n)$, $n\geq2$, where it is easily verified that
$kn+1$ and $n$ are relatively prime. The central charges
and conformal weights are given by
\bea
 c^{(k,n)}&=&1-6\frac{((k-1)n+1)^2}{n(kn+1)}\nn
  &=&1-6\frac{(k-1)^2}{k}-6\frac{(k^2-1)}{k^2n}+{\cal O}(1/n^2)\nn
 \D_{r,s}^{(k,n)}&=&\frac{((kn+1)r-ns)^2-((k-1)n+1)^2}{4n(kn+1)}\nn
  &=&\frac{(kr-s)^2-(k-1)^2}{4k}+\frac{k^2(r^2-1)-(s^2-1)}{4k^2n}
  +{\cal O}(1/n^2)
\label{kn}
\eea
with limits
\bea
 c^{(k)}&=&\lim_{n\rightarrow\infty}c^{(k,n)}\ =\ 1-6\frac{(k-1)^2}{k}\nn
 \D_{r,s}^{(k)}&=&\lim_{n\rightarrow\infty}\D_{r,s}^{(k,n)}\ =\ 
  \frac{(kr-s)^2-(k-1)^2}{4k},\ \ \ \ \ \ \ r,s\in\Z_>
\label{k}
\eea
These are seen to correspond to the similar values in the
(non-minimal) model ${\cal M}(k,1)$ on the boundary of the set of
minimal models.
The model ${\cal M}(1,1)$ is thus related to the limit of the sequence
of {\em unitary} minimal models ${\cal M}(n+1,n)$, and has central
charge $c^{(1)}=1$. The limit of the non-unitary minimal models
${\cal M}(2n+1,n)$ is associated to ${\cal M}(2,1)$ which has appeared 
in the literature
in studies of $c=-2$ LCFT \cite{Kau}. The other integer central charges
obtained in this way are $c^{(3)}=-7$ and $c^{(6)}=-24$.
As we shall discuss presently, Jordan-cell structures can be constructed for
all $k\geq1$.

There is a natural embedding of the Kac table associated to 
${\cal M}(kn_1+1,n_1)$ into the Kac table associated to
${\cal M}(kn_2+1,n_2)$ if $n_1\leq n_2$, mapping 
$\phi_{r,s}^{(k,n_1)}$ to $\phi_{r,s}^{(k,n_2)}$.
Note, however, that the conformal weights and representations
in general will be altered. Our point here is that if
$(r,s)$ is admissible for $n_0$, it will be admissible
for all $n\geq n_0$. We thus have a natural notion of
sequences of primary fields: $(\phi_{r,s}^{(k,n_0)},
\phi_{r,s}^{(k,n_0+1)},...)$.
The parameter $n_0$ is essentially immaterial since we are
concerned with the properties of the sequences as $n\rightarrow\infty$.
We therefore choose to denote a sequence simply as 
$\Upsilon_{r,s}^{(k)}$ (from whose indices the minimal $n_0$ can be
determined anyway). 

These sequences may be organized in equivalence classes, where
$\Upsilon_{r,s}^{(k)}$ and $\Upsilon_{u,v}^{(k)}$ 
are said to be equivalent if they approach the same conformal
weight. For $\Upsilon_{r,s}^{(k)}$ and $\Upsilon_{u,v}^{(k)}$ 
to be equivalent it is required that
$(ku-v)^2=(kr-s)^2$, that is,
\ben
  I:\ \ (u,v)\ =\ (r+q,s+kq),\ \ \ \ \ \ r,s,u,v\in\Z_>,\ \ q\in\Z
\label{I}
\een
or
\ben
 II:\ \ (u,v)\ =\ (-r+q,-s+kq),\ \ \ \ \ \ r,s,u,v\in\Z_>,\ \ q\in\Z
\label{II}
\een
In either case, the approached weight is $\D_{r,s}^{(k)}$ in
(\ref{k}). The equivalence becomes trivial (i.e.,
$\Upsilon_{r,s}^{(k)}=\Upsilon_{u,v}^{(k)}$) if $q=0$ in case $I$
or if $q=2r$ and $s=kr$ in case $II$.

Since our objective is to illustrate the limiting procedure of the
previous section, we should consider two equivalent but different
sequences of primary fields. We have two cases to analyze.
In case $I$, and with reference to the general construction above,
we consider the two primary fields $\var_n^{(k)}=\phi_{r,s}^{(k,n)}$
and $\psi_n^{(k)}=\phi_{r+q,s+kq}^{(k,n)}$ (excluding, of course,
the trivial case $q=0$) in the minimal model ${\cal M}(kn+1,n)$. We find that
\bea
 a_n^{(k)}-b_n^{(k)}&=&\frac{q(2n(s-rk)-2r-q)}{4n(kn+1)}\nn
 &=&\frac{q(s-rk)}{2kn}-\frac{q(2s+qk)}{4k^2n^2}+{\cal O}(1/n^3)
\label{abI}
\eea
which for large enough $n$ is non-vanishing. The map 
(\ref{lin}) and (\ref{n}) then reads
\ben
 \left(\begin{array}{l} \Phi^{(k)}_n\\ \Psi^{(k)}_n\end{array}\right)\ =\
  \left(\begin{array}{lc} 
    \sqrt{\frac{q(2n(s-rk)-2r-q)A}{4n(kn+1)C_{r,s}^{(k)}}}&
     0\\ \\
   \sqrt{\frac{4n(kn+1)A}{q(2n(s-rk)-2r-q)C_{r,s}^{(k)}}}&
 \sqrt{\frac{q(2n(s-rk)-2r-q)B-4n(kn+1)A}{q(2n(s-rk)-2r-q)C_{r+q,s+kq}^{(k)}}}
    \end{array}\right)
     \left(\begin{array}{l} \phi_{r,s}^{(k,n)}\\ 
    \phi_{r+q,s+kq}^{(k,n)}\end{array}\right)  
\label{linI}
\een
where $C_{r,s}^{(k)}=C_{\phi_{r,s}^{(k,n)}}$ has been chosen
independent of $n$.
The resulting Jordan cell given by $\Phi^{(k)}$ and $\Psi^{(k)}$, defined as
in (\ref{PhiPsi}), has conformal weight $\D_{r,s}^{(k)}$ given in (\ref{k}).

Similarly, in case $II$ we consider $\var_n^{(k)}=\phi_{r,s}^{(k,n)}$
and $\psi_n^{(k)}=\phi_{-r+q,-s+kq}^{(k,n)}$ 
(again excluding the trivial case) and find 
\bea
 a_n^{(k)}-b_n^{(k)}&=&\frac{q(2n(rk-s)+2r-q)}{4n(kn+1)}\nn
 &=&\frac{q(rk-s)}{2kn}+\frac{q(2s-qk)}{4k^2n^2}+{\cal O}(1/n^3)
\label{abII}
\eea
The map becomes
\ben
 \left(\begin{array}{l} \Phi^{(k)}_n\\ \Psi^{(k)}_n\end{array}\right)\ =\
  \left(\begin{array}{lc} 
    \sqrt{\frac{q(2n(rk-s)+2r-q)A}{4n(kn+1)C_{r,s}^{(k)}}}&
     0\\ \\
   \sqrt{\frac{4n(kn+1)A}{q(2n(rk-s)+2r-q)C_{r,s}^{(k)}}}&
 \sqrt{\frac{q(2n(rk-s)+2r-q)B-4n(kn+1)A}{q(2n(rk-s)+2r-q)
  C_{-r+q,-s+kq}^{(k)}}}
    \end{array}\right)
     \left(\begin{array}{l} \phi_{r,s}^{(k,n)}\\ 
    \phi_{-r+q,-s+kq}^{(k,n)}\end{array}\right)  
\label{linII}
\een

The formulas simplify a bit in some cases. For example,
the numerator in (\ref{abI}) factorizes as $(2kn-1)(s'^2-r^2)$
when $s=s'k+2rk$ and $q=-r+s'$, whereas the numerator in (\ref{abII})
factorizes as $(2kn+1)(r^2-s'^2)$ when $s=s'k$ and $q=r+s'$.

We conclude that Jordan cells can be constructed for all conformal weights
in the spectrum of ${\cal M}(k,1)$ given in (\ref{k}).
It has also been found that there are numerous ways of constructing a
Jordan cell of a given weight in that spectrum. We shall comment more
on these issues in the final section. In \cite{Kau} on a LCFT with $c=-2$ and 
spectrum related to ${\cal M}(2,1)$, only a {\em subset} of the conformal 
weights in the full spectrum are associated to Jordan cells. The remaining
values correspond to ordinary primary fields. Our construction above
does not a priori distinguish between these two subsets as all
the weights in the spectrum may be associated to Jordan cells.
It would be interesting to understand the origin of this discrepancy.

\section{Logarithmic limits of SCFT}

\subsection{General construction}

The concept of a Jordan cell carries over to the $N=1$ superconformal
case \cite{KAG} where it may be represented straightforwardly 
in the superspace formalism \cite{MS}. 

Let $\xi=(z,\theta)$ be an $N=1$ superspace coordinate with associated
superderivative $D=\theta\pa+\pa_\theta$. Since $\theta$ is a
Grassmann-odd (anti-commuting) variable, a superfield, $\hat{\phi}$,
expands trivially as
\ben
 \hat{\phi}(\xi)\ =\ \phi(z)+\theta\la(z)
\label{superfield}
\een
A {\em primary} superfield of (super)conformal weight $\D$
may be characterized by
\ben
 \hat{T}(\xi_1)\hat{\phi}(\xi_2)\ =\ \frac{\D\theta_{12}
   \hat{\phi}(\xi_2)}{(\xi_{12})^2}
  +\frac{\left(\theta_{12}\pa_2+\frac{1}{2}D_2\right)
    \hat{\phi}(\xi_2)}{\xi_{12}}
\label{prim}
\een
where $\xi_{12}=\xi_1-\xi_2-\theta_1\theta_2$ 
and $\theta_{12}=\theta_1-\theta_2$. The generator of superconformal
transformations (\ref{prim}), $\hat{T}$, is the odd linear combination
\ben
 \hat{T}(\xi)\ =\ \theta T(z)+\frac{1}{2}G(z)
\label{That}
\een
differing by the factor 1/2 from the convention used in \cite{MS}.
$T$ is the ordinary Virasoro generator with central charge $c$, 
whereas $G$ is its primary spin-3/2 superpartner satisfying
\ben
 G(z)G(w)\ =\ \frac{2c/3}{(z-w)^3}+\frac{2T(w)}{z-w}
\label{GG}
\een
In the expansion (\ref{superfield}) of a primary superfield,
the two fields $\phi$ and $\la$ are primary of weights
$\D$ and $\D+1/2$, respectively, with respect to the Virasoro generator $T$.

A superconformal Jordan cell consists of two superfields:
a primary superfield, $\hat{\Phi}$, of (super)conformal weight $\D$, and
its non-primary partner, $\hat{\Psi}$.
They satisfy
\bea
 \hat{T}(\xi_1)\hat{\Phi}(\xi_2)&=& \frac{\theta_{12}\D
   \hat{\Phi}(\xi_2)}{(\xi_{12})^2}
  +\frac{\left(\theta_{12}\pa_2+\frac{1}{2}D_2\right)
    \hat{\Phi}(\xi_2)}{\xi_{12}}  \nn
 \hat{T}(\xi_1)\hat{\Psi}(\xi_2)&=& \frac{\theta_{12}\left(\D
   \hat{\Psi}(\xi_2)+\hat{\Phi}(\xi_2)\right)}{(\xi_{12})^2}
     +\frac{\left(\theta_{12}\pa_2+\frac{1}{2}D_2\right)
    \hat{\Psi}(\xi_2)}{\xi_{12}}  
\label{supercell}
\eea
The associated two-point functions are of the form
\ben
 \langle\hat{\Phi}(\xi_1)\hat{\Phi}(\xi_2)\rangle\ =\ 0,\ \ \ \ 
 \langle\hat{\Phi}(\xi_1)\hat{\Psi}(\xi_2)\rangle\ =\ 
   \frac{A}{(\xi_{12})^{2\D}},\ \ \ \ 
 \langle\hat{\Psi}(\xi_1)\hat{\Psi}(\xi_2)\rangle\ =\ 
   \frac{B-2A\ln\xi_{12}}{(\xi_{12})^{2\D}}
\label{supertwo}
\een
with structure constants $A$ and $B$. Our goal is to construct
such a system in the limit of a sequence of SCFTs.

The construction is a straightforward extension of the one discussed
in the previous section on ordinary CFT.
We thus consider a sequence of superconformal models $SM_n$, 
$n\in\Z_>$, with central charges, $c_n$, converging to the finite value
\ben
 \lim_{n\rightarrow\infty}c_n\ =\ c
\label{superc}
\een
It is assumed that $SM_n$ contains a pair of primary superfields, 
$\hat{\var}_n$ and
$\hat{\psi}_n$, of weights $\D+a_n$ and $\D+b_n$, respectively,
satisfying (\ref{ab}).
Their two-point functions are assumed to be of the form
\bea
 \langle\hat{\var}_n(\xi_1)\hat{\var}_n(\xi_2)\rangle&=&
   \frac{C_{\hat{\var}}}{(\xi_{12})^{2(\D+a_n)}}\nn
 \langle\hat{\psi}_n(\xi_1)\hat{\psi}_n(\xi_2)\rangle&=&
   \frac{C_{\hat{\psi}}}{(\xi_{12})^{2(\D+b_n)}}\nn
  \langle\hat{\var}_n(\xi_1)\hat{\psi}_n(\xi_2)\rangle&=&0
\label{superphivar}
\eea
We now introduce the linear and invertible map
\ben
 \left(\begin{array}{ll} \hat{\Phi}_n\\ \hat{\Psi}_n\end{array}\right)\ =\
  \left(\begin{array}{ll} \al_n&\beta_n\\ \g_n&\delta_n
    \end{array}\right)
     \left(\begin{array}{ll} \hat{\var}_n\\ \hat{\psi}_n\end{array}\right)  
\label{superlin}
\een
and define the superfields
\ben
 \hat{\Phi}:=\ \lim_{n\rightarrow\infty}\hat{\Phi}_n,\ \ \ \ \ \ \ 
 \hat{\Psi}:=\ \lim_{n\rightarrow\infty}\hat{\Psi}_n
\label{superPhiPsi}
\een
It turns out that a map similar to (\ref{n}) also applies
in this case:
\bea
 \al_n&=&\sqrt{\frac{(a_n-b_n)A}{C_{\hat{\var}}}},\ \ \ \ \ \ \ \ 
  \beta_n\ =\ 0\nn
 \g_n&=&\sqrt{\frac{A}{(a_n-b_n)C_{\hat{\var}}}},\ \ \ \ \ \ \ \delta_n\ =\ 
  \sqrt{\frac{(a_n-b_n)B-A}{(a_n-b_n)C_{\hat{\psi}}}}
\label{supern}
\eea
as it is straightforward to show that the superfields defined
in (\ref{superPhiPsi}) indeed constitute a 
superconformal Jordan cell of  weight $\D$.

\subsection{Logarithmic limits of superconformal minimal models}

A superconformal minimal model, ${\cal SM}(p,p')$,
is characterized by two integers whose difference is
even, with $(p-p')/2$ and $p$ (or equivalently $p'$) coprime.
Without loss of generality, we are here following the convention
that $p\geq p'+2$ and $p'\geq2$.
The central charge is given by
\ben
 c\ =\ \frac{3}{2}-3\frac{(p-p')^2}{pp'}
\label{cs}
\een
whereas the primary fields have conformal weights
\ben
 \D_{r,s}\ =\ \frac{(rp-sp')^2-(p-p')^2}{8pp'}+\frac{1}{32}(1-(-1)^{r+s}), 
  \ \ \ \ \ \ \ \ 1\leq r<p', \ \ 1\leq s< p
\label{Dmms}
\een
These fields are subject to the same field identification 
as in (\ref{iden}), and for $p$ and $p'$ both even, the
field $\phi_{p'/2,p/2}$ is unaffected by this identification.
The superconformal minimal model ${\cal SM}(p,p')$ is
unitary provided $p=p'+2$.

The Neveu-Schwarz (NS) sector contains the fields with
$r+s$ even, while fields with $r+s$ odd belong to the
Ramond sector. A primary field of weight $\D_{r,s}$ 
in the NS sector has
a superpartner of weight $\D_{r,s}+1/2$ 
together with which it makes up a primary superfield
of weight $\D_{r,s}$. Since the general construction outlined above
is based on superfields, 
we shall eventually treat the two sectors separately.
 
For each positive integer $k$ we consider the sequence of superconformal
minimal models ${\cal SM}((2k-1)n+2,n)$, $n\geq2$. It is easily verified that
the difference $\{(2k-1)n+2\}-n$ is even, and that half of this difference is
relatively prime to $n$. The central charges
and conformal weights are given by
\bea
 c^{(k,n)}&=&\frac{3}{2}-
   3\frac{\left((2k-1)n+2-n\right)^2}{\left((2k-1)n+2\right)n}\nn
  &=&\frac{3}{2}-\frac{12(k-1)^2}{2k-1}-\frac{24k(k-1)}{(2k-1)^2n}+
     {\cal O}(1/n^2)\nn
  \D^{(k,n)}_{r,s}&=&\frac{\left(r((2k-1)n+2)
     -sn\right)^2-\left((2k-1)n+2-n\right)^2}{8((2k-1)n+2)n}
  +\frac{1}{32}(1-(-1)^{r+s})\nn
   &=&\frac{(r(2k-1)-s)^2-4(k-1)^2}{8(2k-1)}+\frac{1}{32}(1-(-1)^{r+s})\nn
     &+&\frac{(2k-1)^2(r^2-1)-(s^2-1)}{4(2k-1)^2n}+ {\cal O}(1/n^2)
\label{superkn}
\eea
with limits
\bea
 c^{(k)}&=&\lim_{n\rightarrow\infty}c^{(k,n)}\ =\ 
  \frac{3}{2}-\frac{12(k-1)^2}{2k-1}\nn
  \D^{(k)}_{r,s}&=&\lim_{n\rightarrow\infty}\D^{(k,n)}_{r,s}\nn
  &=& \frac{(r(2k-1)-s)^2-4(k-1)^2}{8(2k-1)}+\frac{1}{32}(1-(-1)^{r+s}), 
  \ \ \ \ \ \ \ \  r,s\in\Z_>
\label{superk}
\eea
These are seen to correspond to the similar values in the 
non-minimal model ${\cal SM}(2k-1,1)$ on the boundary of the set of
minimal models.
The model ${\cal SM}(1,1)$ is thus related to the limit of the sequence
of {\em unitary} minimal models ${\cal SM}(n+2,n)$, and has central
charge $c^{(1)}=3/2$. The limit of the non-unitary minimal models
${\cal SM}(3n+2,n)$, corresponding to $k=2$, is associated to 
${\cal SM}(3,1)$ with central
charge $c=-5/2$. These are the only two models of this kind
with half-integer central charge. 

Motivated by the construction of Jordan cells in conformal minimal models,
we now consider when two sequences of primary (super)fields,
$\Upsilon_{r,s}^{(k)}$ and $\Upsilon_{u,v}^{(k)}$,
associated to the sequence ${\cal SM}((2k-1)n+2,n)$ are equivalent.
First, it is observed that a sequence in the NS sector ($r+s$ even)
cannot approach the same conformal weight as a sequence in the
Ramond sector ($u+v$ odd). 
The condition for equivalence in either sector then reads
$((2k-1)r-s)^2=((2k-1)u-v)^2$, that is,
\ben
  I:\ \ (u,v)\ =\ (r+q,s+(2k-1)q),\ \ \ \ \ \ r,s,u,v\in\Z_>,\ \ q\in\Z
\label{superI}
\een
or
\ben
 II:\ \ (u,v)\ =\ (-r+q,-s+(2k-1)q),\ \ \ \ \ \ r,s,u,v\in\Z_>,\ \ q\in\Z
\label{superII}
\een
with identity for $q=0$ in case $I$ and 
for $q=2r$ and $s=(2k-1)r$ in case $II$.

Superconformal Jordan cells may now be constructed straightforwardly
by combining (\ref{supern}) with the prescription in the conformal case above. 
We thus work out $a^{(k)}_n-b^{(k)}_n$ in both cases (\ref{superI})
and (\ref{superII}), and the superconformal Jordan cells emerge in the 
limit (\ref{superPhiPsi}).

Regarding the Ramond sector, we propose to deal with it
in the same way as
we dealt with primary fields in the previous section on ordinary CFT.
Again, we work out $a^{(k)}_n-b^{(k)}_n$ in both cases (\ref{superI})
and (\ref{superII}), and the conformal Jordan cells emerge as $n$ approaches infinity.

Due to the similarity between the two sectors, we may present the
results in a unified way. In the case (\ref{superI}) we find
\bea
 \left(\begin{array}{l} \Phi^{(k)}_n\\ \Psi^{(k)}_n\end{array}\right)&=&
  \left(\begin{array}{lc} 
    \sqrt{\frac{q(n(s-r(2k-1))-2r-q)A}{2n((2k-1)n+2)C_{r,s}^{(k)}}}&
     0\\ \\
   \sqrt{\frac{2n((2k-1)n+2)A}{q(n(s-r(2k-1))-2r-q)C_{r,s}^{(k)}}}&
 \sqrt{\frac{q(n(s-r(2k-1))-2r-q)B-2n((2k-1)n+2)A}{q(n(s-r(2k-1))-2r-q)
    C_{r+q,s+(2k-1)q}^{(k)}}}
    \end{array}\right)\nn\nn
  &\times&
     \left(\begin{array}{l} \phi_{r,s}^{(k,n)}\\ 
    \phi_{r+q,s+(2k-1)q}^{(k,n)}\end{array}\right)  
\label{superlinI}
\eea
while in the case (\ref{superII}) we have
\bea
 \left(\begin{array}{l} \Phi^{(k)}_n\\ \Psi^{(k)}_n\end{array}\right)&=&
  \left(\begin{array}{lc} 
    \sqrt{\frac{q(n(r(2k-1)-s)+2r-q)A}{2n((2k-1)n+2)C_{r,s}^{(k)}}}&
     0\\ \\
   \sqrt{\frac{2n((2k-1)n+2)A}{q(n(r(2k-1)-s)+2r-q)C_{r,s}^{(k)}}}&
 \sqrt{\frac{q(n(r(2k-1)-s)+2r-q)B-2n((2k-1)n+2)A}{q(n(r(2k-1)-s)+2r-q)
  C_{-r+q,-s+(2k-1)q}^{(k)}}}
    \end{array}\right)\nn\nn
  &\times&
     \left(\begin{array}{l} \phi_{r,s}^{(k,n)}\\ 
    \phi_{-r+q,-s+(2k-1)q}^{(k,n)}\end{array}\right)  
\label{superlinII}
\eea
In both cases, $C_{r,s}^{(k)}=C_{\phi_{r,s}^{(k,n)}}$ has been chosen
independent of $n$. Also, in the NS sector
the two maps (\ref{superlinI}) and (\ref{superlinII})
correspond to the even parts of the superfields
but are identical to the maps for the superfields themselves. The
maps for the superfields are
therefore not written explicitly.

\section{Conclusion}

We have discussed how certain limits of sequences of CFTs and SCFTs may 
correspond to logarithmic CFTs and SCFTs.
Particular emphasis has been put on minimal
models, and we have found that certain logarithmic limits 
may be associated to non-minimal (super)conformal 
models on the boundary of the set of (super)conformal minimal models.
An infinite family of such logarithmic limits has been 
proposed in the ordinary as well as in the superconformal case.

The map (\ref{lin}) (and (\ref{superlin}) in the superconformal case)
will in general only affect a small subset of the full
spectrum of fields. One should therefore expect that 
different linear and invertible maps of the complete
set of fields
\ben
 \left(\begin{array}{c}\mbox{}\\ {\rm new}\\ {\rm set}\\
  \mbox{}\end{array}\right)\ =\ \left(\begin{array}{c}\mbox{}\\
 ({\rm singular})\\ {\rm matrix}\\ \mbox{}\end{array}\right)
 \left(\begin{array}{c}\mbox{}\\ {\rm old}\\ {\rm set}\\
  \mbox{}\end{array}\right)
\label{genmap}
\een
in general would result in inequivalent models in the
limit $n\rightarrow\infty$. In particular, maps that become
singular in this limit
seem to be proned to alter the spectrum.
The naive limit of the unitary
series ${\cal M}(n+1,n)$, where the map is governed by the
identity matrix, is related to the discussion in \cite{RuW,RoW}
where it has been found
to corespond to a non-rational but non-logarithmic CFT
with $c=1$.

Alternatives to the sequences ${\cal M}(kn+1,n)$ and 
${\cal SM}((2k-1)n+2,n)$ can, of course, be envisaged.
Following our general construction of (super)conformal
Jordan cells, these would potentially yield different
logarithmic models in the appropriate limits.
A possible classification of the models thus obtainable
is an interesting problem to pursue.
It should be noted that the map (\ref{genmap}), followed by
a limiting procedure, in many cases will lead to a non-logarithmic model.
The resulting model is most likely non-rational, though.

Here we confine ourselves to indicating how one may construct
Jordan cells of any weight associated to
the general model ${\cal M}(p,p')$ where $p$ and $p'$ are coprime.
To this end, let $\Nat_{p'}$ denote the set of positive integers relatively 
prime to $p'$, and consider
\ben
 {\cal M}(pn+p'^a,p'n)
\label{M}
\een 
for some positive integer $a$.
It follows that $pn+p'^a$ 
and $p'n$ are relatively prime for $n\in\Nat_{p'}$. 
For $p'\neq1$, one could replace $p'^a$ with any non-trivial
product of non-negative powers of the prime factors of $p'$, and
(\ref{M}) would still correspond to a minimal model.
For fixed $a$, we now consider the sequence of models (\ref{M}) where 
$n\in S_{p'}$ and $S_{p'}$ is an infinite subset of $\Nat_{p'}$.
As $n$ increases and approaches infinity, the central charge and
spectrum of conformal weights will approach the similar values
in the model ${\cal M}(p,p')$. Our prime example above
corresponds to $p'=1$ and $S_{p'}=\Nat_{\geq2}$,
while another class of examples was pointed out to us by
A. Nichols and is given by the minimal models
${\cal M}(p^2n,pp'n+1)$ which
approach ${\cal M}(p,p')$ in the limit $n\rightarrow\infty$.
These examples are obviously not the only possibilities for
obtaining the central charge and spectrum of conformal
weights of ${\cal M}(p,p')$, in a limiting procedure.

Our construction pertains to (super)conformal Jordan cells of rank two.
We have recently found \cite{Ras}, though, that it extends to Jordan 
cells of rank $r$ \cite{RAK} where
\ben
 T(z)\Psi_{(j)}(w)\ =\ 
 \frac{\D\Psi_{(j)}(w)+(1-\delta_{j,0})\Psi_{(j-1)}(w)}{(z-w)^2}
 +\frac{\pa_w\Psi_{(j)}(w)}{z-w},\ \ \ \ \ \ \ j=0,1,...,r-1
\label{r}
\een
Here $\Psi_{(0)}$ is a primary field while the other $r-1$ fields
are not.

We would like to comment on the many ways 
Jordan cells may be obtained according to our outline, cf. 
the observation following (\ref{linII}), in particular.
This suggests that there could be
infinitely many fields or Jordan cells of a given weight.
One attempt to circumvent this, if so desired, is to consider
(\ref{genmap}) by
supplementing the maps (\ref{lin}) by scaling the 
'unwanted' fields by $n$ (or perhaps even higher-degree polynomials
in $n$ or $\sqrt{n}$) to prevent them from
showing up in operator products of the 'wanted' fields.
This could potentially result in finitely many fields of a given
weight.
The finer structure of the operator-product algebra may reveal, though, 
that it is impossible to avoid the unwanted fields, but this issue
is beyond the scope of the present work.
Field identifications may alternatively resolve the problem.
It is emphasized that we are dealing with chiral fields only.
As the issue of locality in LCFT appears much more subtle and 
complicated than in ordinary CFT, attempting to construct the
full LCFT may also put severe and unexpected constraints
on the permitted chiral structure.

Now, the possibility of infinitely many
rank-two Jordan cells appearing with a given conformal weight
could perhaps be understood as infinitely many ways of
extracting rank-two Jordan cells from a single Jordan cell
of {\em infinite} rank. Such a structure could possibly help
explaining why some logarithmic models have been
found (see \cite{Flo-9605}) to display features similar 
to rational CFTs, despite the highly
non-rational nature of LCFT.

It is known that the (super)conformal minimal models can
be represented in terms of cosets of affine current algebras.
The unitary series, in particular, can be given by coset constructions
with diagonal embeddings \cite{GKO}:
\bea
 {\cal M}(n+1,n)&\simeq&\frac{\widehat{su}(2)_{n-2}\oplus
   \widehat{su}(2)_1}{\widehat{su}(2)_{n-1}},\ \ \ \ \ \ \ \ \ n\geq2\nn
  {\cal SM}(n+2,n)&\simeq&\frac{\widehat{su}(2)_{n-2}\oplus
   \widehat{su}(2)_2}{\widehat{su}(2)_{n}},\ \ \ \ \ \ \ \ \ n\geq2
\label{cosets}
\eea
The levels of the constituent affine Lie algebras are indicated
by subindices, and are all integer.
The coset constructions of all other series of (super)conformal
minimal models will involve non-integer, though fractional levels.
The integer nature of the levels in (\ref{cosets}) allows
one to describe the cosets in terms of gauged Wess-Zumino-Witten
models. This in turn seems to suggest that the $c=1$ logarithmic CFT and
the $c=3/2$ logarithmic SCFT discussed above
may admit geometric interpretations obtained as limits
of the geometries associated to the unitary series.
This could potentially mimic the Penrose limits known from
studies of (super)gravity solutions and properties of space-time, see
\cite{SS} and references therein.
We hope to address elsewhere this exciting possibility along with 
the other open questions and speculations above.
\vskip.5cm
\noindent{\em Acknowledgements}
\vskip.1cm
\noindent The author thanks C. Cummins, P. Mathieu, A. Nichols,
M. Walton and, in particular, M. Flohr for comments.

\end{document}